\documentclass[review,times,12pt]{elsarticle}
\usepackage{graphicx}
\usepackage{color}
\usepackage{bm}
\usepackage{units,nicefrac}
\DeclareGraphicsExtensions{.eps}
\journal{Journal of Non-Crystalline Solids}

\usepackage{lineno}

\begin{document}
	\begin{frontmatter}
		\title{Atomic bonding in equilibrium single-component melts. The cases of arsenic, antimony and bismuth}
		
		\author[kfu,urfc]{Artem A. Tsygankov\corref{cor1}}
		\cortext[cor1]{Corresponding author}
		\ead{tsigankov.artiom@yandex.ru}
		
		\author[kfu,ufrc]{Bulat N. Galimzyanov}
		\ead{bulatgnmail@gmail.com}
		
		\author[kfu,ufrc]{Anatolii V. Mokshin}
		\ead{anatolii.mokshin@mail.ru}
		
		\affiliation[kfu]{organization={Kazan Federal University},
			addressline={Kremlevskaya 16},
			city={Kazan},
			postcode={420008},
			state={Tatarstan Republic},
			country={Russia}}
		\affiliation[ufrc]{organization={Udmurt Federal Research Center of the Ural Branch of RAS},
			addressline={Tatyana Baramzina 34}, 
			city={Izhevsk},
			postcode={426067}, 
			state={Udmurtia Republic},
			country={Russia}}
		
		\begin{abstract}
			In liquid pnictogens, quasi-stable structures can be formed near melting temperature. The nature of their stability does not have the unified point of view. In the present work, the task of determining the degree of atomic bonding in these structures is solved using the Crystal Orbital Hamilton Population (COHP) method.
			The original results of \textit{ab-initio} simulation of arsenic, antimony and bismuth melts near their melting temperatures are used. It is shown that the features of the electron interaction at the level of $p$-orbitals determine the characteristic bond lengths and angles between atoms. It has been established that the stability of structures decreases according to a power law with an increase in the atomic mass of a chemical element and the number of atoms in the structure. The obtained results clarify the understanding the mechanisms of formation of quasi-stable structures in pnictogen melts from first principles.
		\end{abstract}
		
		\begin{highlights}
			\item Formation of quasi-stable structures is primarily due to \textit{p}-orbital interactions
			\item Contribution of \textit{s}-orbitals to the formation of these structures can be neglected
			\item Stability of these structures decrease as power-law with increasing atomic number
			\item Quasi-stable structures have a maximum size related to their constituent elements
			\item Exceeding this maximum size can lead to structural instability or collapse
		\end{highlights}
		
	\end{frontmatter}
	\newpage
	\section{Introduction}
\begin{figure*}[h!]
		\centering
		\includegraphics[width=1.0\linewidth]{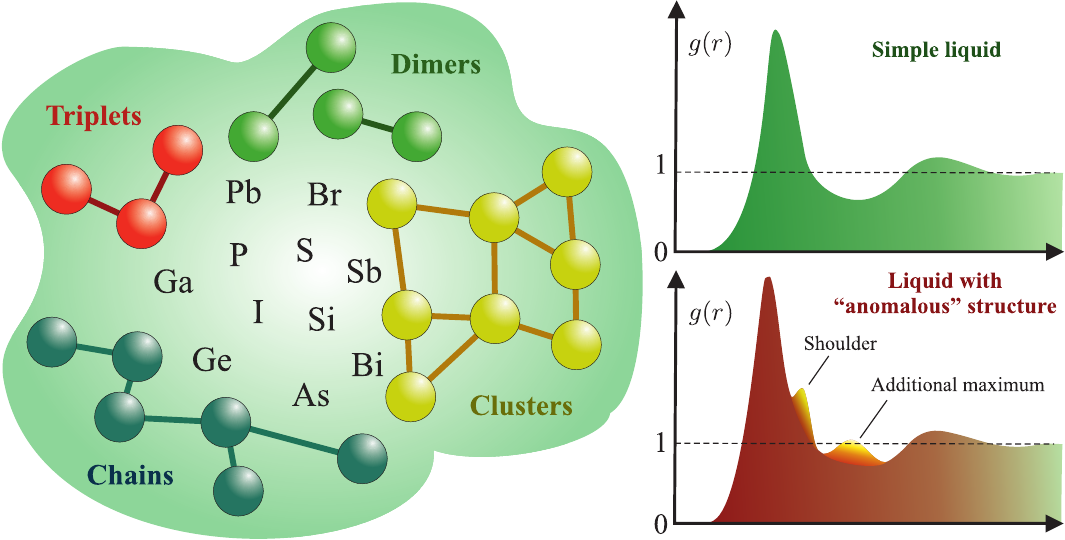}
		\caption{Schematic representation of quasi-stable structural formations that can exist in the melts of pure pnictogens. The radial distribution function of atoms for a simple liquid without an "anomalous" structure and for a liquid with quasi-stable structures. The curves are schematic and presented for clarity.}
		\label{fig_1}
	\end{figure*}

	In several polyvalent monoatomic melts near the melting temperature, an specific "short-range" order is observed, which may be attributed to the presence of small-sized structures with finite lifetimes. The existence of these structures has been confirmed through neutron and X-ray diffraction experiments as well as \textit{ab-initio} calculations~\cite{Flores-Ruiz_2022_liquid_s, Shi_2024_structures_in_sn_sb, Tsygankov_liquid_sb_2024, Xu_2022, Yang_2021_p, Yang_2024_liquid_s}. Namely, the presence of quasi-stable structures in the melt leads to a distortion of the static structure factor $S(k)$ and the radial distribution function $g(r)$. This distortion is observed as an asymmetric main peak or additional shoulders/broadening in functions $S(k)$ and $g(r)$. Such structural anomalies are exist in Ge, Si, Sb, Bi, Pb, I, Br and S melts at the pressure $1$~atm. Such the structures also exist in P and As melts under high pressure (see Figure~\ref{fig_1})~\cite{Xu_2022, Mokshin_liquid_ga_2020, Umnov_1993, Brazhkin_1997, Zou_1987, Scopigno_2007_liquid_s, Sastry_2003, Principi_2005}. These melts exhibit bound atomic pairs (dimers) as well as structures as clusters. For example, molecular-like tetrahedral configurations have been identified in P and As melts~\cite{Yang_2021_p, Li_1990_liquid_as_theor, Bernal_1960}, whereas Sb and Bi melts predominantly form dimers or small clusters~\cite{Akola_2014_clusters_bi, Jones_2017_liquid_sb}.
	
	Several ideas have been proposed to explain the origin of such structures in polyvalent melts and to interpret the corresponding experimental and molecular dynamics simulation results. A notable example is the double hard-sphere model that was developed for Ge, Sb, Ga, Sn and Bi melts. This model explains the formation of quasi-stable structures by the presence in the melts of atoms with two different effective diameters~\cite{Orton_liquid_bi_1979}. According to this model, the structure factor can be decomposed into two components that correspond to different states -- structured and disordered (liquid). Although this approach provides qualitative insight into atomic configurations within the melt, it inadequately describes the local environment observed in most melts~\cite{Xiao_commentary_1989}. Furthermore, alternative interpretations suggest analyzing the liquid structure as a distorted crystalline phase, particularly for melts of arsenic and bismuth~\cite{Hafner_liquid_as_1989, Kawakita_Kikuchi_2018}. The argument for this phenomenon comes from determining interatomic distances and coordination numbers, which are close to the values of crystalline phase~\cite{Gaspard_1988_peierls_transititon}. Notably, several researches has demonstrated that the formation of quasi-stable structures in certain melts of elements from the 13--16 groups of the Periodic Table can be explained by Friedel oscillations in the effective interatomic potential~\cite{Jank_1990_divalent_elements, Hafner_periodic_table_1989}. The characteristic wavelength of these oscillations is $2k_f$, where $k_f$ -- Fermi wave vector. It is noteworthy that the quantity $2k_f$ is universal for polyvalent monatomic melts in groups from $13$ to $16$.
	
	It should be emphasized that there is still no unified understanding of the structural properties of polyvalent melts. One such case is liquid Ga, where experimental and computational methods lead to different results. On the one hand, \textit{ab-initio} simulations and nuclear magnetic resonance (NMR) frequency shift measurements revealed the presence of short-lived Ga-Ga dimers, which are remnants of the $\alpha$-Ga crystalline phase~\cite{Gong_1993_liquid_ga_dimers}. The concentration of such atomic pairs was estimated to be about 6\%, depending on the thermodynamic conditions~\cite{Xiong_2017_liquid_ga}. On the other hand, the results of resent studies are found no evidence for the presence of atomic pairs in Ga melts~\cite{Mokshin_liquid_ga_2020}. These conclusions are based on the close-packed quasi-soft sphere model, which assumes that the short-range order in Ga melt is represented by a range of correlation lengths. This model reproduces the experimentally observed structural features of single-component melts in the absence of quasi-molecular units. This was confirmed by \textit{ab-initio} simulations of liquid Ga~\cite{Mokshin_liquid_ga_2020, Mokshin_liquid_ga_2015}.
	
	Moreover, in some cases complex formations may be present, as was shown for liquid sulfur. It can form both rings (most often from 8 atoms $S_8$) and structures in the form of chains~\cite{Flores-Ruiz_2022_liquid_s, Vahvaselka_1988_liquid_s}. At the same time, the chains formation mechanism remained unclear. One of the recent studies explained this process using the Bader charge analysis method, which is usually used to analyze molecules and crystals~\cite{Yang_2024_liquid_s}. Namely, the proposed polymerization mechanism is a multi-stage process. Thermal fluctuations induce opening of ring $S_8$, where one of the bonds brakes. This process generates charge polarization at the edge atoms of the ring, making them active. At this step, the opened ring of sulfur atoms can join another closed guest ring. In this scenario, electron charges are redistributed at the edges of oligomer. New active centers are formed and polymerization spreads further~\cite{Yang_2024_liquid_s}. This result allows one to argue for the application of the analysis methods for molecules to quasi-stable structures in the liquid phase.
	
	In the absence of a unified view on the existence of quasi-stable structures in polyvalent metal melts, it is necessary to estimate the stability and size of these structures through the characteristic energies of interatomic interactions. In the present work, to solve this task the methodology for analyzing the population distributions of projected crystalline orbitals of the Hamiltonian based on quantum-chemical calculations from first principles was used. This approach focuses on equilibrium As, Sb and Bi melts near their melting temperatures, where quasi-stable structures were observed experimentally.

	\section{Simulation details and applied methods}
	Investigation of As, Sb and Bi melts is performed using Vienna Ab-initio Simulation Package (VASP) with ultrasoft pseudopotential~\cite{Kresse_1993,Kresse_1996_1,Kresse_1996_2, Kresse_1994_us}. In the simulation, the NVT ensemble with a fixed density corresponding to the experimental density (pressure) is used. Density is equal to $4.22\cdot10^{-2}~\AA^{-3}$, $3.20\cdot10^{-2}~\AA^{-3}$, $2.89\cdot10^{-2}~\AA^{-3}$ for As, Sb and Bi correspondingly~\cite{Hafner_liquid_as_1989, Waseda_1980}. The considered temperatures are near their melting points. The ratios between the temperatures $T$ and the melting points $T_m$ are equal $T/T_m$ $\approx$ 1.01 for As, $T/T_m$ $\approx$ 1.02 for Sb and $T/T_m$ $\approx$ 1.05 for Bi. Here, melting temperatures are equal 1090 K for As (at the pressure 28 atm), 904 K for Sb and 545 K for Bi. Time step in all simulations was $1$~fs. The number of atoms $N$ is $N=384$ for As, $N=384$ for Sb and $N=432$ for Bi. Thermodynamic conditions (temperature and pressure) corresponding to the experimental conditions were considered: $T=1100$~K and $p=28$~atm for As; $T=923$~K and $p=1$~atm for Sb; $T=573$~K and $p=1$~atm for Bi. The melting temperatures of As, Sb and Bi at this pressures are $T_{m}=1090$~K, $T_{m}=904$~K and $T_{m}=545$~K, respectively. Note that As actively sublimes at high temperatures and ambient pressure. Therefore, empirical and simulation studies of As were carried out at the pressure $p=28$~atm, where sublimation was not observed. 
	
\begin{figure}[h!]
		\centering
		\includegraphics[width=0.9\linewidth]{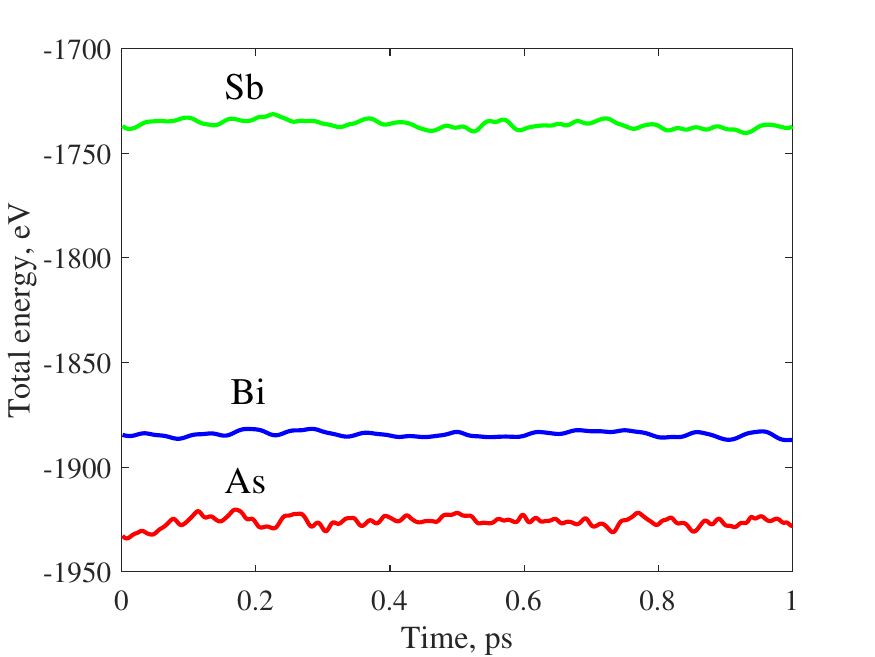}
		\caption{Dependence of the total energy on the simulation time for As, Sb and Bi melts. The energy fluctuates around a constant value, which confirms that the melts are in a thermodynamic equilibrium state.}
		\label{fig_2}
	\end{figure}

	The considered systems are in the thermodynamically equilibrium state. This is confirmed by the fact that the calculated total energy for each system oscillates around a constant value. This is demonstrated at the Figure~\ref{fig_2}, where the dependence of the total energy on the simulation time for As, Sb and Bi is shown. These liquids were also relaxed for 1 ps (1000 time steps). It is worth noting that all calculations were carried out at temperatures close to the melting point $T_m$: $T/T_m\approx$ 1.01 for As, $T/T_m\approx$ 1.02 for Sb, $T/T_m\approx$ 1.05 for Bi. Here, the melting temperatures are 1090 K for As (at the pressure 28 atm), 904 K for Sb and 545 K for Bi.
	
	To explain the stability of anomalous structures, contribution of atomic orbitals to the interaction energy between atoms was analyzed. The characteristic binding energy of atoms in arbitrary quasi-stable structures was estimated using projected crystal Hamilton population (pCOHP)~\cite{Deringer_2011_cohp}. The pCOHP shows the contribution of individual atoms (or orbitals) to the total energy of the band structure. It allows studying chemical bonds, nature of interactions and contribution of energy bands in structures and nanoparticles. Information on the contribution of atomic orbitals and binding energies is obtained from \textit{ab-initio} simulations based on electronic structure data estimated using the projected augmented wave (PAW) method. This method is usually used for high-precision calculations such as the density of states or band structure. 
	 
	It should be noted that pCOHP is widely used for the analysis of the electronic structure of crystalline materials. The application of pCOHP to melts requires special consideration due to the absence of a crystal lattice. Statistical postprocessing is applied to a set of local quasi-stable structures such as dimers, triplets, four-, five- and six-atom structures. This set consists of equilibrium configurations of these local structures obtained at different time steps. Averaging these configurations allows one to determine the average interatomic distances for each type of local structure. Then, using the calculated average interatomic distances, the average values of pCOHP and |IpCOHP| are determined. The averaging is performed over 20 ps of simulation time (20000 timesteps), which is sufficient to obtain reliable results. The errors of measurements were calculated using statistical analysis of the results from multiple independent molecular dynamics simulations. The deviation from the most probable results is estimated as a calculation error.
	
	In the presence of quasi-stable structures, the melt may exhibit a certain average order (signs of a crystal structure)~\cite{Kawakita_Kikuchi_2018}. In this case, pCOHP can be used to analyze the contribution of individual atoms or atomic groups to the overall distribution of states~\cite{Yan_Li_2025}. In addition, temperature and pressure significantly affect to melt structure and its dynamics, potentially changing the local arrangement of atoms and electronic spectra. Therefore, thermodynamic conditions must be taken into account when estimating pCOHP.
	
	The following expression is used to estimate the pCOHP~\cite{Deringer_2011_cohp}:
	\begin{equation}\label{eq_pcohp}
		pCOHP_{\mu\nu} (\textbf{k})=\sum_{j} R\left[ P_{\mu\nu j}^{(proj)}(\textbf{k}) H_{\nu\mu}^{(proj)}(\textbf{k}) \right]\times \delta(\epsilon_j(\textbf{k}) - E).
	\end{equation}
	Here $\delta(\epsilon_j(\textbf{k}) - E)$ is the Dirac delta function; $j$ is the index of energy bond of wave function $\left|\psi\right\rangle$, defined from \textit{ab-initio} calculations; $\mu$ is the index of the wave function of an atomic orbital $\phi$ of the first atom; $\nu$ is the index of orbital of the second atom, where the presence of a bond is supposed to be. To calculate pCOHP from the ab-initio simulation data, the PAW method in the Perdew–Burke–Ernzerhof approximation (PAW-PBE) with the kinetic cutoff energy 700 eV was used. This value of cutoff energy is suitable for the performed calculations since it ensures high accuracy of the charge density and wave functions. The $k$-point grid size of 12$\times$12$\times$12 was also used, which is optimal according to Ref.~\cite{Deringer_2011_cohp}. 
	
	In the expression (\ref{eq_pcohp}), the quantity $H_{\mu\nu}^{(proj)}(\textbf{k})$ is the Hamiltonian in the plane wave basis $H^{(pw)}$, expressed as a local basis of the known function $\left|\phi\right\rangle$ with energy of the bond $\epsilon_j(\textbf{k})$:
	\begin{equation}
		H_{\mu\nu}^{(proj)} (\textbf{k}) = \left\langle \phi_{\mu} \left| H^{(pw)} \right| \phi_{\nu}  \right\rangle = 
		\sum_{j} \epsilon_j(\textbf{k})T^*_{j\mu}(\textbf{k}) T_{j\nu}(\textbf{k}).
	\end{equation}
	In the expression (\ref{eq_pcohp}), the quantity $P_{\mu\nu j}^{(proj)}(\textbf{k})$ is the projection density matrix of each energy band $j$ at each point $\textbf{k}$ with elements
	\begin{equation}
		P_{\mu\nu j}^{(proj)}(\textbf{k}) = T^*_{j\mu}(\textbf{k})T_{j\nu}(\textbf{k}),
	\end{equation}
	where $T(\bf{k})$ is the "transitition matrix" with elements $T_{j\mu}(\textbf{k}) = \left\langle \psi_j(\textbf{k})|\phi_\mu \right\rangle$. As a result, if the value of $pCOHP(E)$ is positive, then the contribution to the total interaction energy is antibonding (i.e. atoms show tendency to not form bonds). If the result is negative, then the contribution is bonding, which helps form quasi-stable structures. This technique is implemented in LOBSTER package, which helps in extracting useful information about orbitals and their contributions~\cite{Maintz_2016_lobster}.
	\begin{figure*}[h!]
		\centering
		\includegraphics[width=1\linewidth]{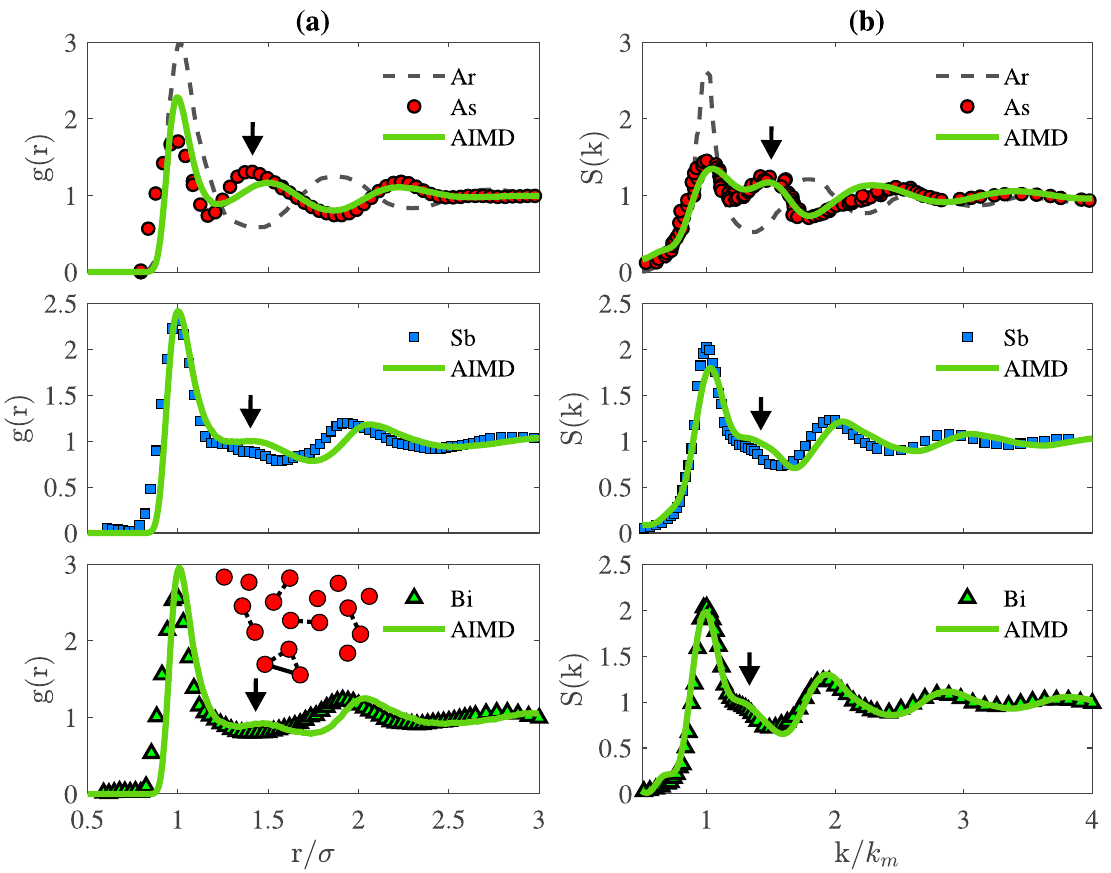}
		\caption{
			(a) Radial distribution function of atoms g(r) and (b) statical structure factor S(k) for As, Sb and Bi melts. The markers indicate experimental X-ray diffraction data taken from~\cite{Hafner_liquid_as_1989} (for As) and~\cite{Waseda_1980}~(for Sb, Bi). The solid green lines show the obtained ab-initio simulation (AIMD) results. In top panels, the functions g(r) and S(k) for liquid argon at the temperature 85 K are presented~\cite{Yarnell_1973_liquid_ar}. The g(r) and S(k) curves were obtained under identical thermodynamic conditions at their phase diagrams. Characteristic lengths for dimers and triplets are shown by arrows. In figure, $\sigma$ is the effective diameter of an atom (2.55~\AA~for As, 3.3~$\AA$ for Sb and 3.4~$\AA$ for Bi), $k_m$ is the position of first maximum in the function S(k) (2.32~$\AA^{-1}$ for As, 2.15~$\AA^{-1}$ for Sb and 2.15~$\AA^{-1}$ for Bi).}
		\label{fig_3}
	\end{figure*}
	
	\section{Results and discussions}
	
	According to the results of empirical and simulation studies, in As, Sb and Bi melts atoms can form dimers and triplets~\cite{Tsygankov_liquid_sb_2024, Li_1990_liquid_as_theor, Galimzyanov_liquid_bi_2023}. This is confirmed by calculating the lifetimes of these structures and by the presence of characteristic bond lengths and angles. For example, in Sb and Bi melts, the elementary units of quasi-stable structures are dimers and triplets with bond angles of $45^\circ$ and $90^\circ$. Their characteristic lengths are $3.07$~\AA~and $4.8$~\AA~in the case of Sb and $3.25$~\AA~and $4.7$~\AA~for Bi~\cite{Tsygankov_liquid_sb_2024, Galimzyanov_liquid_bi_2023}. In the case of As, the presence of tetrahedral structures was shown with bond length $2.5$~\AA~and bond angle $90^\circ$~\cite{Li_1990_liquid_as_theor}. 
	
	Figure~\ref{fig_3} shows the \textit{ab-initio} simulation results and experimental X-ray scattering data obtained for As, Sb and Bi~\cite{Hafner_liquid_as_1989, Waseda_1980}. The simulation results are in agreement with experimental data – the structural anomaly is reproduced and correctly describes the shoulder in the functions $S(k)$ and $g(r)$. The nature of interaction of atoms in As, Sb and Bi melts is not symmetrical and is characterized by directionality, which generates additional shoulders/peaks in the functions g(r) and S(k). In structure factor, the shoulders appear in wave number intervals [2.74; 3.96]~$\AA^{-1}$ for As, [2.6; 3.0]~$\AA^{-1}$ for Sb and [2.6; 3.2]~$\AA^{-1}$ for Bi. Additional peaks in the radial distribution function appear at distances [2.97; 4.75]~\AA~for As, [3.7; 4.7]~\AA~for Sb and [3.8; 5.2]~\AA~for Bi. These characteristic lengths correspond to interatomic distances that arise only when local structures with two and three bonded atoms (i.e. dimers and triplets) are formed~\cite{Tsygankov_liquid_sb_2024, Galimzyanov_liquid_bi_2023}. For comparison, Figure~\ref{fig_3} (upper panels) shows the results for liquid Ar obtained by neutron scattering~\cite{Yarnell_1973_liquid_ar}. This system is a typical example of a simple liquid in which there are no structural anomalies. It contains only peaks corresponding to the first and second coordinations. The absence of anomalies in the structure of liquid argon is due to the fact that the nature of the interaction of atoms is spherically symmetric.
	
\begin{figure*}[ht]
		\centering
		\includegraphics[width=1.0\linewidth]{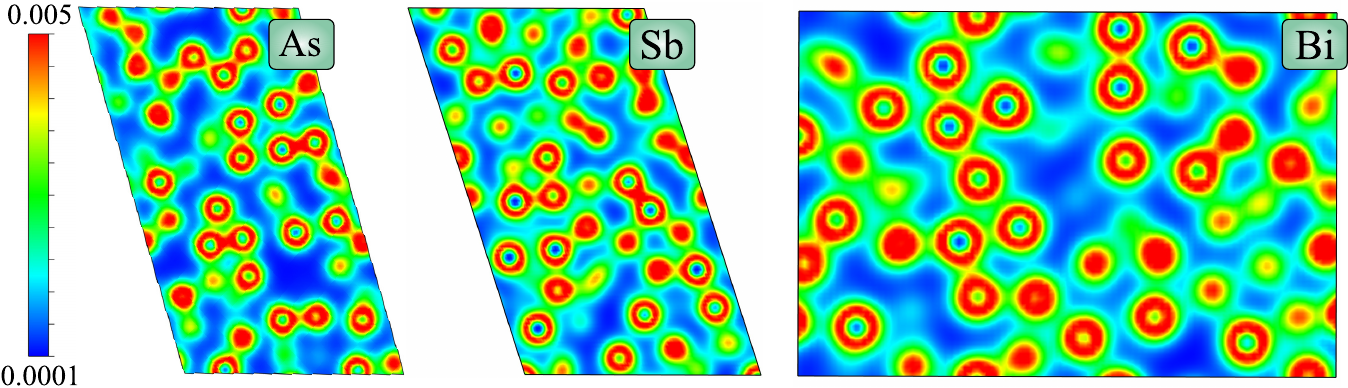}
		\caption{Snapshots of charge density distribution for As, Sb and Bi. In the case of As and Sb, the simulation cell is rhombohedral, while the Bi atoms are located in a rectangular cell. This explains the difference in the shape of the resulting snapshots. The color indicates the change in charge density. The maximum density, which has the value $0.005e$, corresponds to red. The minimum charge density with the value $0.0001e$ corresponds to dark blue color.
		} \label{fig_4}
	\end{figure*}

	Based on the results of \textit{ab-initio} calculations, instantaneous snapshots of the charge density distribution near the melting point were obtained. The snapshots were cut by the Miller indices (0 1 2) of the simulation cell. As shown in Figure~\ref{fig_4}, electrons are strongly localized in the regions of formation of dimers and triplets. In regions of high electron localization, the charge density value reaches $0.005e$ ($e$ is the modulus of the elementary charge of an electron), while in regions with lower density, the value drops to $0.0001e$.
	
	The energy characteristic that determines the degree of bonding of atoms in quasi-stable structures is calculated by integrating the pCOHP function to the Fermi level $E_f$, called IpCOHP~\cite{Maintz_2013}:
 
	\begin{equation} \label{eq_ipcohp}
		IpCOHP = \int_{-\infty}^{E_f}pCOHP(E)~dE.
	\end{equation}
	The IpCOHP calculation was performed for dimers, triplets and larger clusters of four, five and six bonded atoms. In this case it is usually recommended to consider only the $s$- and $p$-orbitals for the pCOHP calculations~\cite{Deringer_2011_cohp, Dronskowski_1993}. Figures~\ref{fig_4}(a)--\ref{fig_4}(f) show that below the Fermi level the $\vert$-pCOHP(E)$\vert$ projections for dimers and triplets are similar in the As, Sb and Bi systems. However, their intensity decreases with increasing atomic mass. This trend is also reflected in the calculated values IpCOHP (see Table~\ref{table_1}). 

\begin{figure*}[ht]
		\centering
		\includegraphics[width=1.0\linewidth]{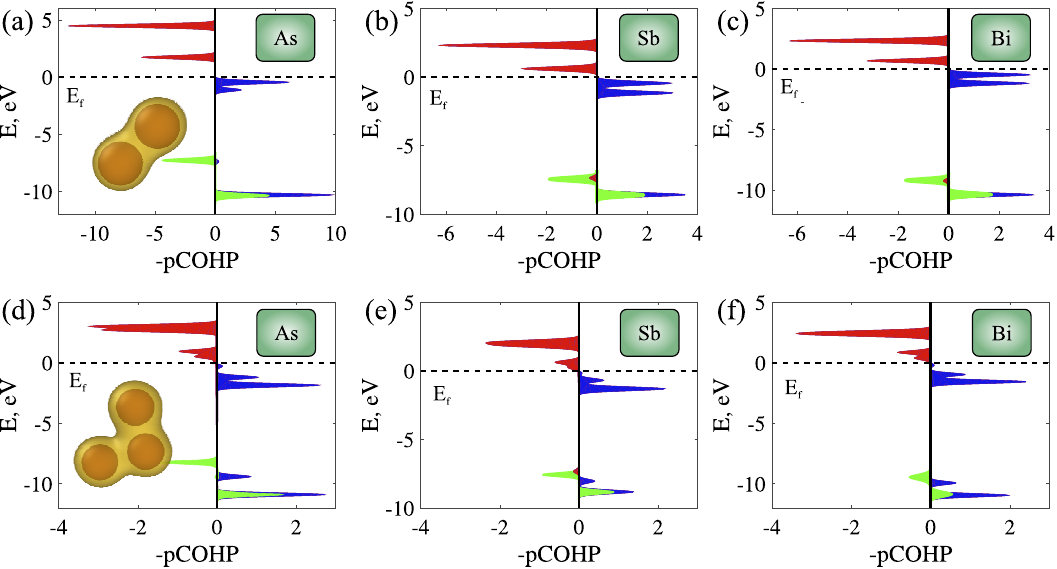}
		\caption{The projected crystal orbital Hamilton populations (pCOHP) for dimers (a)--(c) and triplets (d)--(f) obtained for As, Sb and Bi melts near melting temperatures. Energies are presented relative to the Fermi energy $E_f$ and shifted to zero. Negative (i.e., bonding) contributions are plotted on the right and shaded blue, while positive (i.e., antibonding) contributions are plotted on the left and shaded red. Green zone shows results for calculation of pCOHP with $s$-orbital. Its intergral (IpCOHP) is equal to zero.}\label{fig_5}
	\end{figure*}

	The results show that the $s$-orbital does not contribute to the interatomic bonding, as evidenced by the zero values of IpCOHP for $s$-electrons [Figure~\ref{fig_5}]. At the same time, the integral IpCOHP is not zero when considering $s$- and $p$-orbitals. Thus, the nature of the interaction of atoms in quasi-stable structures depends on the features of the $p$-orbital structure, which has the shape of three orthogonal dumbbells.
	
	\begin{figure}[ht]
		\centering
		\includegraphics[width=1.0\linewidth]{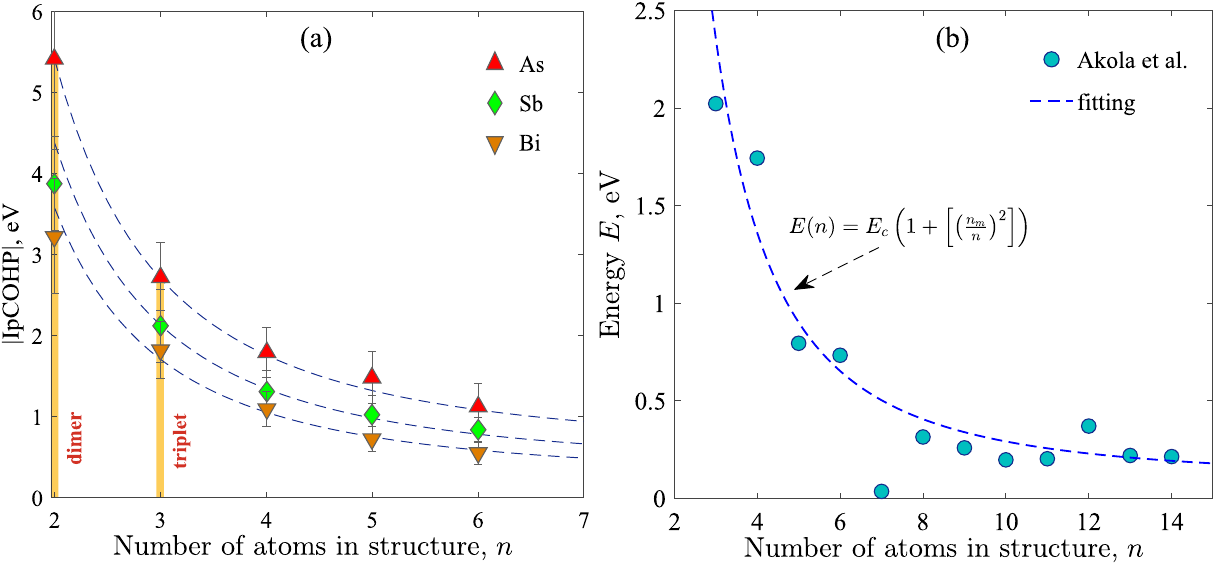}
		\caption{(a) The value of $\vert$IpCOHP$\vert$ as a function of the number of atoms in a quasi-stable structure. The markers show the results of \textit{ab-initio} calculations. (b) Average energy of a quasi-stable structure as a function of the number of atoms in this structure. These data are taken from Ref.~\cite{Akola_2014_clusters_bi}. The dotted curve is the result of Eq.~(\ref{eq_cohpSize}).} \label{fig_6}
	\end{figure}
	
	The found values of IpCOHP are negative. This means that the interaction in quasi-stable structures is bonding. With increasing number of electrons and atomic radius, the value of IpCOHP decreases. For comparison, for the As-As dimer, the value of IpCOHP is $-5.41\pm1.1$~eV, and for the As-As-As triplet, where the arsenic atom has atomic number $33$ and the outer electron shell 4s$^2$4p$^3$. In the case of antimony with atomic number 51 as the outer electronic shell 5s$^2$5p$^3$, the value of IpCOHP is $-3.87\pm0.5$~eV for Sb-Sb and $-2.12\pm0.2$~eV for Sb-Sb-Sb [see Table~\ref{table_1}]. In the case of bismuth with atomic number $83$, having the electron configuration 6s$^2$6p$^3$, IpCOHP is $-3.22\pm0.7$~eV for Bi-Bi and $-1.82\pm0.3$~eV for Bi-Bi-Bi. Thus, IpCOHP for dimers is approximately twice as large (in absolute value) as for triplets, indicating greater stability. Therefore, dimers are more stable than triplets.
	
	The value of the integral $\vert$IpCOHP$\vert$ decreases with the increase of the number of atoms in the structure, as shown in Figure~\ref{fig_6}(a). This decrease in value indicates a decrease in the stability of the structure. As can be seen from Figure~\ref{fig_6}(a), $\vert$IpCOHP$\vert$ decreases according to a power law depending on the number of atoms $n$ in a quasi-stable structure:
	\begin{equation} \label{eq_cohpSize}
		|IpCOHP|(n) = \frac{ |IpCOHP|_{c}}{2}\left[1+\left(\frac{n_m}{n}\right)^{2}\right],\,n=2,3,4....
	\end{equation}
	Here, $n_{m}$ is the maximum average number of atoms, where structure is capable of maintaining stability. According to Eq.~(\ref{eq_cohpSize}), quasi-stable structures with sizes up to $n_m=6$ and $n_m=7$ are the most common for As and Sb. In the case of Bi melt, the size of structures can reach $n_m=8$ atoms [see Table~\ref{table_2}]. The maximum size of quasi-stable structures increases in proportion to increasing charge number of atoms. If a structure contains more bound atoms than the maximum value, the structure will decay into separate fragments due to thermal motion of atoms. The value of $\vert$IpCOHP$\vert$$_c$ in Eq.~(\ref{eq_cohpSize}) corresponds to the equilibrium crystalline phases As-I, Sb-I and Bi-I at the pressure 1 atm. For these crystalline phases, values of $\vert$IpCOHP$\vert$$_{c}$ vary in the range [$0.39$, $0.75$]. Early studies are demonstrated that quasi-stable structures formed in melts can ``inherit'' the structural parameters of the crystalline phase~\cite{Yang_2024_liquid_s, Akola_2014_clusters_bi, Jones_2017_liquid_sb, Kawakita_Kikuchi_2018, Gaspard_1988_peierls_transititon}. Therefore, it can be assumed that the value of $\vert$IpCOHP$\vert$$_{c}$ characterizes the degree of "inheritance" of crystalline phase: the higher the value of $n$, the more similar parameters of quasi-stable structure are to nearest crystalline phase at an isobar.
	
	In addition to the data on the dependence of $\vert$IpCOHP$\vert$  on the size of quasi-stable structures, it is possible to estimate the behavior of the average energy of the quasi-stable structure on its size (Figure~\ref{fig_6}(b)). As can be seen from Figure~\ref{fig_6}(b), the dependence of the average energy on the size of structure taken from~\cite{Akola_2014_clusters_bi} is described by a similar functional dependence as for $\vert$IpCOHP$\vert$ (as Eq.~\ref{eq_cohpSize}). In this case, the coefficients of the equation are $E_c = 0.09$ eV and $n_m = 15$. It is also important to note that an increase in the number of atoms decreases the average energy of the structure, i.e. stabilizes it. Therefore, the structure of a polyvalent monatomic liquid with anomalous structural properties in group 15 can be considered in terms of two competing phenomena. The first is the interaction of atoms with each other, tending to form quasi-stable structures at the corresponding isobar. The second is the thermal motion of atoms that destroys quasi-stable structures. Their presence explains the appearance of quasi-stable structures, including the ``locally preferred structures'' introduced by H. Tanaka~\cite{Tanaka_2020_LL_polyamrph}.
	
	\section{Conclusion}
	
	The formation of quasi-stable structures in the form of dimers and triplets in As, Sb and Bi melts is caused, first of all, by the structure of the outer electron shell of these atoms. It is shown that the COHP method is applicable for an approximate characteristic of the degree of bonding between atoms in these structures based on \textit{ab-initio} simulations. From the results obtained by the COHP method, it follows that the characteristic correlation lengths and valence angles observed in the experiments are consequence of the interaction of atoms at the $p$-orbital level. An expression in the form of a power function is obtained, which relates the COHP integral to the number of atoms in a quasi-stable structure. Using this expression, the average limiting size of the structures is estimated. It is shown that the degree of bonding of atoms in quasi-stable structures decreases with increasing size of this structure and with increasing mass number of the atom. On the other hand, thermal motion is one of the main factors limiting the size of the structures. Thus, in the work~\cite{Greenberg_2009_EPL}, using the example of Bi melt, it was shown that quasi-stable structures are not formed at temperatures above $1013$\,K. This transition temperature is significantly lower than the boiling point of Bi, which is equal to $\approx1840$\,K. It can be assumed that for this melt there may be an equilibrium phase containing quasi-stable structures isolated from the usual equilibrium liquid phase. In the diagram, these two phases are probably separated by a liquid-liquid transition line, where a change in density is observed, as was shown in the case of phosphorus~\cite{Yang_2021_p}. The results of this work do not take into account the possibility of more complex phase diagrams of these melts. Detailed studies are needed not only near the melting line, but also in the region of a possible liquid-liquid transition.
	
	\section*{Acknowledgment} 
	\noindent The work was carried out on the basis of the grant provided by
	the Academy of Sciences of the Republic of Tatarstan (Kazan, Russia) in 2024 for the implementation of fundamental and applied research work in scientific and educational organizations, enterprises and organizations of the real sector of the economy of the Republic of Tatarstan.

	\newpage
	\begin{table}[h]
		\begin{center}
			
			\begin{tabular}{cccccc}
				\hline
				& IpCOHP$^{(2)}$ & IpCOHP$^{(3)}$ & IpCOHP$^{(4)}$ & IpCOHP$^{(5)}$ & IpCOHP$^{(6)}$  \\
				\hline
				As & $-5.41\pm1.1$ & $-2.72\pm0.3$ & $-1.79\pm0.25$ & $-1.48\pm0.2$ & $-1.12\pm0.15$  \\
				Sb & $-3.87\pm0.5$ & $-2.12\pm0.2$ & $-1.31\pm0.15$ & $-1.02\pm0.14$ & $-0.83\pm0.12$  \\
				Bi & $-3.22\pm0.7$ & $-1.82\pm0.3$ & $-1.09\pm0.16$ & $-0.72\pm0.15$ & $-0.55\pm0.14$  \\
				\hline
			\end{tabular}
			\caption{Average values of IpCOHP (eV) for dimers, triplets, and for quasi-stable structures consisting four, five and six atoms.} 
			\label{table_1}
		\end{center}
	\end{table}	
	
	\begin{table}[h]
		\begin{center}
			
			\begin{tabular}{ccc}
				\hline
				& $\vert$IpCOHP$\vert$$_{c}$, eV & $n_{m}$ \\
				\hline
				As & $0.75\pm0.08$ & $6\pm1$ \\
				Sb & $0.68\pm0.06$ & $7\pm1$ \\
				Bi & $0.39\pm0.05$ & $8\pm1$ \\
				\hline
			\end{tabular}
			\caption{Parameters $\vert$IpCOHP$\vert$$_{c}$ and $n_{m}$ of Eq.~(\ref{eq_cohpSize}).} 
			\label{table_2}
		\end{center}
	\end{table}	
	
\end{document}